\documentclass[conference]{IEEEtran}
\usepackage{amsmath,subfigure}
\usepackage{graphicx}
\usepackage{grffile}
\usepackage{algpseudocode}
\usepackage{algorithm}
\usepackage{epstopdf}
\usepackage{amsmath}
\usepackage{amssymb}

\newcommand{\qed}{\nobreak \ifvmode \relax \else
      \ifdim\lastskip<1.5em \hskip-\lastskip
      \hskip1.5em plus0em minus0.5em \fi \nobreak
      \vrule height0.75em width0.5em depth0.25em\fi}

\begin{document}
\title{A Signal Constellation for Pilotless Communications Over Wiener Phase Noise Channels}

\author{
\IEEEauthorblockN{Shachar Shayovitz and Dan Raphaeli} \\
\vspace{-0.3cm}
\IEEEauthorblockA{Dept. of EE-Systems, Tel-Aviv University \\
Tel-Aviv 69978, Israel\\
Email: shachars@post.tau.ac.il,danr@eng.tau.ac.il}}

\vspace{-0.3cm}
\maketitle

Many satellite communication systems operating today employ low cost upconverters or downconverters which create phase noise. This noise can severely limit the information rate of the system and pose a serious challenge for the detection systems. Moreover, simple solutions for phase noise tracking such as PLL either require low phase noise or otherwise require many pilot symbols which reduce the effective data rate.

In the last decade we have witnessed many research papers on the design of signal constellations for channels with phase noise. In \cite{farbod2012}, the phase noise is assumed to be a memoryless Tikhonov distributed stochastic process. The authors compute the maximal information rate assuming an i.u.d input and find the optimal constellation which achieves this rate. The resulting constellations for different Tikhonov distributions are all asymmetrical constellations and look very different from the APSK/MPSK constellations, usually used in wireless communications today. However, the physical phase noise process is not white, the most widely used stochastic model for phase noise is the Wiener process \cite{Hajimiri1998}. The memoryless Tikhonov distributed phase noise is obtained as the residual error after a PLL is already locked on the signal. Thus, whenever the phase noise is too strong for allowing the use of PLL, the phase noise is not memoryless, and the methods devised in \cite{farbod2012} cannot be used. For the best of our knowledge, there is no result for an optimal discrete constellation, in terms of maximal information rate, for the Wiener phase noise channel.

Combining strong error correcting code like LDPC or turbo code with phase noise mitigation algorithm requires iterative algorithm which iterates between the symbol demodulator and the decoder. These algorithms rely on the insertion of pilot symbols for solving phase ambiguities and bootstrapping the decoding process. Alternatively, noncoherent methods are used, usually with some degradation in performance \cite{peleg2000iterative},\cite{barb2011}. However, the insertion of pilots sequence, reduces the effective information rate. Thus a tradeoff exists, the more pilots are inserted, the better the estimation process is, but the effective information rate is lower. In addition, in some existing standards, like DVB-S2, the pilots are too separated between each other to allow good tracking in high phase noise.

In order to increase the effective information rate, we propose a signal constellation which does not require pilots, \textbf{at all}, in order to converge in the decoding process. In this contribution, we will present a signal constellation which does not require pilot sequences, but we require a signal that does not present rotational symmetry. For example a simple MPSK cannot be used.Moreover, we will provide a method to analyze the proposed constellations and provide a figure of merit for their performance when iterative decoding algorithms are used.

\section{System Model}
We consider the transmission of a sequence of complex modulation symbols $\mathbf{c} = (c_{0},c_{1},...,c_{K-1})$
over an AWGN channel affected by carrier phase noise. The discrete-time baseband complex equivalent channel model at the receiver is given by:
\begin{equation}\label{sys_model}
    r_{k} = c_{k}e^{j\theta_{k}}+n_{k} \;\;\;\;  k=0,1,...,K-1.
\end{equation}
We assume a Wiener process phase noise stochastic model:
\begin{equation}\label{Wiener}
    \theta_{k} = \theta_{k-1} + \Delta_{k}
\end{equation}
where ${\Delta_{k}}$ is a real, i.i.d gaussian sequence with $\Delta_{k} \sim \textsl{N}(0,\sigma_{\Delta}^{2})$.

\begin{figure}
  \centering
  \includegraphics[width=7.5cm]{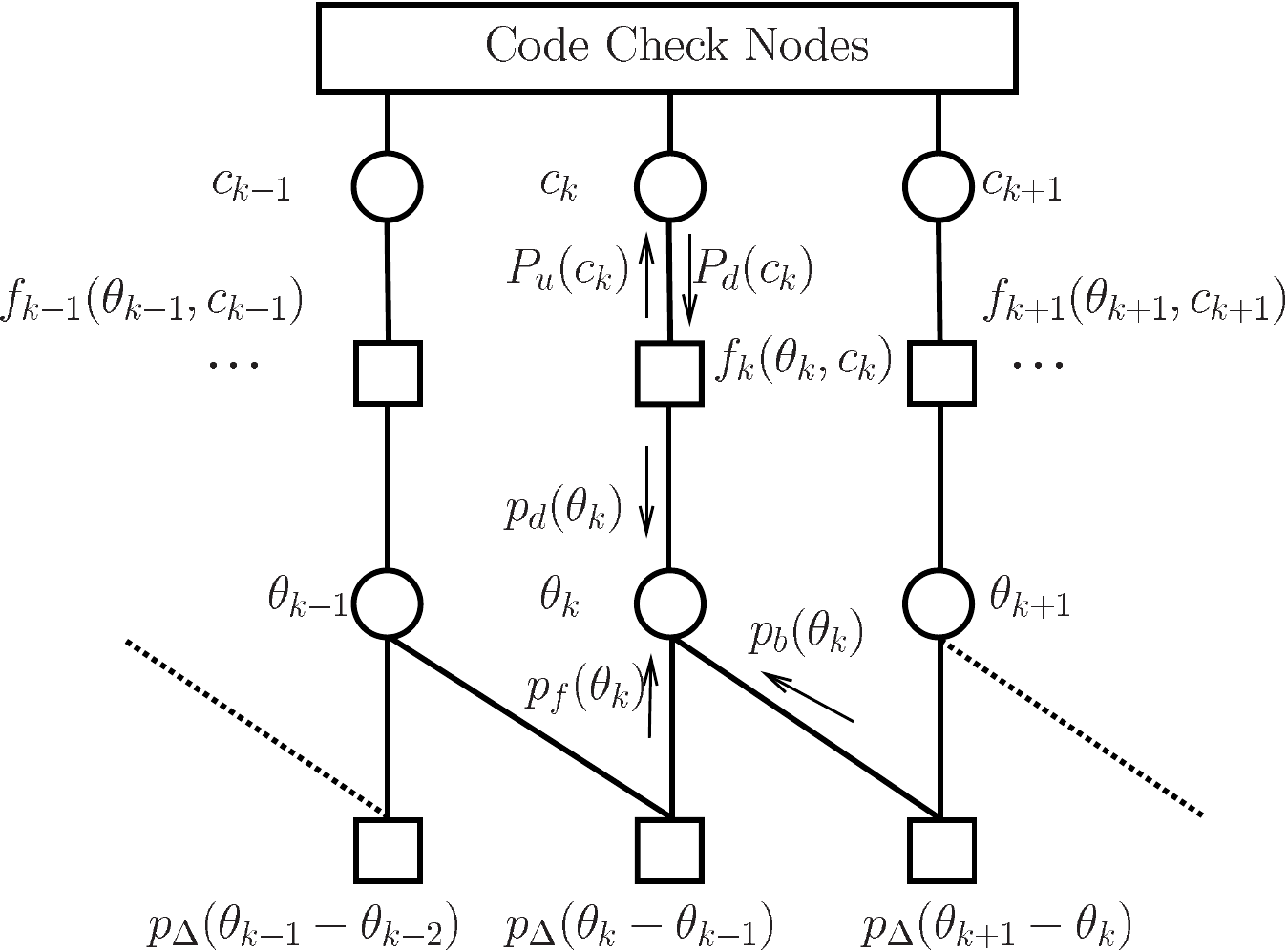}\\
  \caption{Factor graph representation of the joint posterior distribution}\label{fig:fg}
\end{figure}

We are interested in computing a MAP decoder for the data symbols. This decoder can be designed using the Sum and Product Algorithm on the factor graph representation of the joint posterior distribution which was given in \cite{barb2005} and is shown in Fig. \ref{fig:fg}.

The resulting Sum \& Product messages are:
\begin{equation}\label{pf}
    p_{f}(\theta_{k}) \propto \int_{0}^{2\pi}p_{f}(\theta_{k-1})p_{d}(\theta_{k-1})p_{\Delta}(\theta_{k}-\theta_{k-1})d\theta_{k-1}
\end{equation}
\begin{equation}\label{pb}
    p_{b}(\theta_{k}) \propto \int_{0}^{2\pi}p_{b}(\theta_{k+1})p_{d}(\theta_{k+1})p_{\Delta}(\theta_{k+1}-\theta_{k})d\theta_{k+1}
\end{equation}
\begin{equation}\label{pd}
    p_{d}(\theta_{k}) \propto \sum_{m=0}^{M-1} P_{d}(c_{k}=e^{j\frac{2\pi m}{M}}) f_{k}(c_{k},\theta_{k})
\end{equation}
\begin{equation}\label{Pu}
    P_{u}(c_{k}) \propto \int_{0}^{2\pi}p_{f}(\theta_{k})p_{b}(\theta_{k})f_{k}(c_{k},\theta_{k})d\theta_{k}
\end{equation}

where,
\begin{equation}\label{fk}
    f_{k}(c_{k},\theta_{k}) \propto \exp\{-\frac{|r_{k}-c_{k}e^{j\theta_{k}}|^{2}}{2\sigma^{2}}\}
\end{equation}

\begin{equation}\label{p_del}
    p_{\Delta}(\theta_{k}) = \sum^{\infty}_{l=-\infty}g(0,\sigma_{\Delta}^{2},\theta_{k}-l2\pi)
\end{equation}

and $M$,$r_{k}$,$P_{d}$, $\sigma^{2}$ and $g(0,\sigma_{\Delta}^{2},\theta)$ are the constellation order, received base band signal, symbol soft information from LDPC decoder, AWGN variance and Gaussian distribution respectively.

Since a direct implementation of the above messages is impractical due to the fact that the phase is a continuous random variable, several approximation algorithms have been proposed. In \cite{barb2005}, section 3, an algorithm which quantizes the phase noise and performs an approximation of the sum \& product algorithm (SPA) is presented. This algorithm (called DP - discrete phase in this paper), can approach the BER performance of the MAP decoder if the quantization level is large enough. However, this algorithm requires large computational resources to reach high accuracy, rendering it not practical for some real world applications.

In \cite{shachar2012}, a low complexity algorithm was proposed which does not quantize the phase process, but approximates the SPA messages according to the following Tikhonov mixtures,

\begin{equation}\label{new_pf}
    p_{f}(\theta_{k}) = \sum_{i=1}^{N_{f}}\alpha^{k}_{i}f^{f}_{i}(\theta_{k})
\end{equation}

\begin{equation}\label{new_pb}
    p_{b}(\theta_{k}) = \sum_{i=1}^{N_{b}}\beta^{k}_{i}f^{b}_{i}(\theta_{k})
\end{equation}

where:
\begin{equation}\label{f_f_i}
f^{f}_{i}(\theta_{k}) =  \frac{e^{Re[z^{f,k}_{i}e^{-j\theta_{k}}]}}{2\pi I_{0}(|z^{f,k}_{i}|)}
\end{equation}

\begin{equation}\label{f_b_i}
f^{b}_{i}(\theta_{k}) =  \frac{e^{Re[z^{b,k}_{i}e^{-j\theta_{k}}]}}{2\pi I_{0}(|z^{b,k}_{i}|)}
\end{equation}

Using mixture approximations in bayesian inference methods, such as SPA, results in an increase of the mixture order from symbol to symbol. In \cite{shachar2012}, there is a mixture reduction algorithm which keeps the mixture order small while keeping the BER levels low. For large enough mixture order, this algorithm can perform as well as the DP algorithm, while having much lower computational complexity. Moreover, this algorithm gives us an insight to the underlying physics of phase noise tracking.

\begin{figure}
  \centering
  \includegraphics[width=7.5cm]{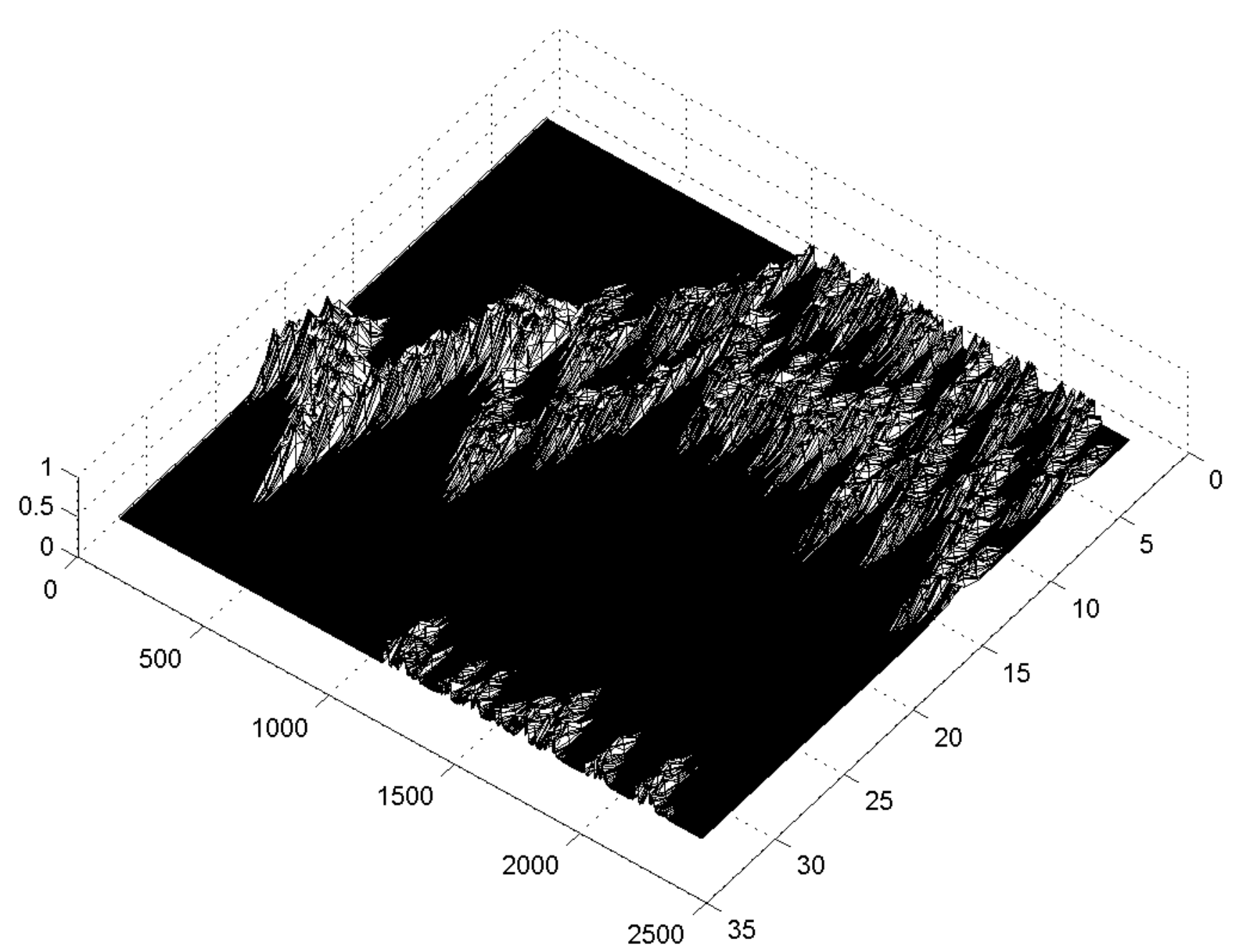}\\
  \caption{SP Phase Noise Forward Messages}\label{fig:splits}
\end{figure}

As shown in Fig. (\ref{fig:splits}), the phase noise messages can be viewed as multiple separate phase trajectories. The mixture algorithm proposed in \cite{shachar2012}, can be viewed as a scheme to map the different mixture components in the phase messages to different phase trajectories and the respective probabilities $\alpha^{k}_{i}$ and $\beta^{k}_{i}$ are the probabilities of each phase noise trajectory. In \cite{shachar2012}, the mixture algorithm receives a mixture describing the next step of all the trajectories and assigns a tracking loop to each trajectory, thus we are able to accurately track all the hypotheses for all the phase trajectories.

\section{Skewed-MPSK and other asymmetrical constellations}
In this section we propose a new signal constellation for pilotless communications over the Wiener phase noise channel. The skewed MPSK (SM) is just one example to Asymmetrical Constellation (AC) that can be used, and the following analysis applies to any constellation. An AC is any constellation which has no rotational symmetry, i.e. there is no $\theta$ such that $c_{k}e^{j\theta}$ is a valid symbol, $\forall k$. As discussed earlier, using symmetrical signal constellations without pilot sequences will result in decoding failures. Suppose we use an MPSK constellation and approximate the forward message well as (\ref{new_pf}). The next step in the SPA is to insert this approximation to (\ref{pf}) and get a bigger mixture representing the next forward message. However, since we use MPSK, then $p_{d}(\theta_{k})$ is simply a mixture of identical Tikhonov distributions, and therefore each mixture component in (\ref{new_pf}) will be multiplied by the same function, resulting with many phase trajectories which behave the same but are separated by multiples of $\frac{2\pi}{M}$, similarly to a cycle slip in a PLL. Therefore, in a pilotless scenario, one must use asymmetrical signal constellations which enable the decoding algorithm to discard certain phase trajectories.

We now define the SM signal constellation,

\begin{equation}\label{SkewMpsk}
  s_{m} = e^{j\frac{2\pi m}{M+skew}}  \;\;\;\;  m=0,1,...,M-1.
\end{equation}

This constellation takes the standard MPSK constellation and instead of having the symbols equally spaced with angle $\frac{2\pi}{M}$ between them, the angles are now not equally spaced thus creating an asymmetry. This of course has some effect on the symbol error probability, but since we are using a long LDPC code the effect should be small for small skew values.

When applying this constellation to the DP algorithm or the mixture model algorithm, this constellation, creates an asymmetrical $p_{d}(\theta_{k})$. Therefore, when using (\ref{pf}), the probabilities ,$\alpha^{k}_{i}$, of the incorrect trajectories in (\ref{new_pf}) are attenuated and decay over time. Therefore, there is a process of eliminating the incorrect phase trajectories in (\ref{new_pf}) and basically lowering the mixture order. If the skew value is high then there is more asymmetry and the decay of the probabilities is faster. This decay process is different for each skew value and can be used to assess the performance of the signal constellation in a tracking scenario.

\subsection{Decay Factor}
We propose a method to analyze the performance of AC over channels corrupted by Wiener phase noise. Without the loss of generality, we will show the results for the forward recursion of the SPA, but the same applies for the backward recursion. Since we can view the phase estimation process as multiple phase trajectories tracking, then we can create a recursion equation for the probabilities, $\alpha^{k}_{i}$, of each trajectory in the forward message (\ref{new_pf}). As discussed earlier, for asymmetrical constellations, the process $\alpha^{k}_{i}$ of a wrong trajectory, decays with $k$. We will show that the expectation process, i.e $\mathbb{E}(\alpha^{k}_{i})$ decays exponentially with $k$. Therefore, the exponential rate, which we denote as decay rate can perform as a figure of merit for the performance of the AC.

The skew value in the SM presents a tradeoff. If the decay rate is large, then the probabilities $\alpha^{k}_{i}$ decay fast and the phase estimation algorithm estimates the correct trajectory fast without the need for pilots. However, if the skew value is too large then the minimum distance is lower and that can affect the code performance. It is easy to assess the degradation in performance by calculating the information rate over the AWGN of such constellation.

We will now compute the recursion equation of the probabilities $\alpha^{k}_{i}$. For the sake of simplicity, we will only address the case of two trajectories, one which is the correct phase noise trajectory and another which resulted due to a phase ambiguity $\phi$, which is a multiple of the symbol spacing , $\frac{2\pi}{M+skew}$.

Assuming that the forward message can be modeled as (\ref{new_pf}),

\begin{equation}\label{decay_pf}
    p_{f}(\theta_{k}) = \sum_{i=1}^{2}\alpha^{k}_{i}f^{f}_{i}(\theta_{k})
\end{equation}

where:
\begin{equation}\label{f_f_1}
f^{f}_{1}(\theta_{k}) =  \frac{e^{Re[z^{f,k}e^{-j\theta_{k}}]}}{2\pi I_{0}(|z^{f,k}|)}
\end{equation}

\begin{equation}\label{f_f_2}
f^{f}_{2}(\theta_{k}) =  \frac{e^{Re[z^{f,k}e^{j\phi}e^{-j\theta_{k}}]}}{2\pi I_{0}(|z^{f,k}|)}
\end{equation}

then applying (\ref{pf}), will result in a new mixture which according to the mixture reduction algorithm in \cite{shachar2012}, can be reduced to a mixture of smaller size. Moreover, we assume that all Tikhonov components cluster according to the $f^{f}_{i}(\theta_{k})$ which created them, thus creating (\ref{mix_tikh_11}), which has only two strong Tikhonov components.

\begin{equation}\label{mix_tikh_11}
p_{f}(\theta_{k+1}) = \sum_{i=1}^{2}\alpha^{k+1}_{i}\frac{e^{Re[\frac{z^{f,k+1}_{i}}{1+\sigma^{2}_{\Delta}|z^{f,k+1}_{i}|}e^{-j\theta_{k+1}}]}}{2\pi I_{0}(|\frac{z^{f,k+1}_{i}}{1+\sigma^{2}_{\Delta}|z^{f,k+1}_{i}|}|)}
\end{equation}

\begin{equation}\label{Z1}
    z^{f,k+1}_{1} = z^{f,k}+\frac{r_{k}\xi^{*}}{\sigma^2}
\end{equation}
\begin{equation}\label{Z2}
    z^{f,k+1}_{2} = z^{f,k}e^{j\phi}+\frac{r_{k}\zeta^{*}}{\sigma^2}
\end{equation}
\begin{equation}\label{coeff1}
    \alpha^{k+1}_{1} = \alpha^{k}_{1}\sum_{l=1}^{M}[P_{d}(c_{k}=e^{j\frac{2\pi l}{M+skew}})\frac{I_{0}(|z^{f,k}+\frac{r_{k}e^{-j\frac{2\pi l}{M+skew}}}{\sigma^2}|)}{I_{0}(|z^{f,k}|)}]
\end{equation}
\begin{equation}\label{coeff2}
    \alpha^{k+1}_{2} = \alpha^{k}_{2}\sum_{l=1}^{M}[P_{d}(c_{k}=e^{j\frac{2\pi l}{M+skew}})\frac{I_{0}(|z^{f,k}e^{j\phi}+\frac{r_{k}e^{-j\frac{2\pi l}{M+skew}}}{\sigma^2}|)}{I_{0}(|z^{f,k}|)}]
\end{equation}

where, $\xi$ and $\zeta$ are symbols taken from the SM constellation.

Using some algebra, we find the recursion equation for the probability of the wrong trajectory (\ref{coeff2}),

\begin{equation}\label{decay}
  \alpha^{k+1}_{2} = \frac{1}{1+\frac{\alpha^{0}_{1}}{\alpha^{0}_{2}}\prod_{i=1}^{k+1}\frac{I_{0}(|\tilde{Z}^{f,i}_{1}|)}{I_{0}(|\tilde{Z}^{f,i}_{2}|)}}
\end{equation}

We assume that approximately the expected value of the probability of the incorrect trajectory decays exponentially with a decay factor $\delta$,

\[\mathbb{E}(\alpha^{k}_{2}) \approx e^{\delta k}\]

Thus, the decay factor $\delta$, can be computed as,
\begin{equation}\label{decay_factor}
  \delta = -\mathbb{E}[\log(\frac{I_{0}(|\tilde{Z}^{i}_{1}|)}{I_{0}(|\tilde{Z}^{i}_{2}|)})]
\end{equation}

Assuming (\ref{Z1}) is correctly tracking the phase noise trajectory $\theta_{k}$, and using (\ref{sys_model}), we can write the following,

\begin{equation}\label{Z1_better1}
    e^{j\theta_{k+1}}(C+\frac{(c_{k}+\tilde{n}_{k})e^{-j\frac{2\pi l}{M+skew}}}{\sigma^2})
\end{equation}

where $C$ and $\tilde{n}_{k}$ are a real number indicating the MSE in estimating the phase noise using the received samples and a complex random variable with the same pdf as $n_{k}$, respectively.

Using (\ref{Z1_better1}), we can numerically compute (\ref{decay1}) and use this decay factor as a figure of merit for the performance of the constellation,

\begin{equation}\label{decay1}
    \delta = -\mathbb{E}[\log(\frac{I_{0}(|C+\frac{(c_{k}+n_{k})e^{-j\frac{2\pi l}{M+skew}}}{\sigma^2}|)}{I_{0}(|Ce^{j\phi}+\frac{(c_{k}+n_{k})e^{-j\frac{2\pi l}{M+skew}}}{\sigma^2}|)})]
\end{equation}

Moreover, note that the proposed analysis method can be applied to to any signal constellation and assess the decay factor.

\begin{figure}
  \centering
  \includegraphics[width=7.5cm]{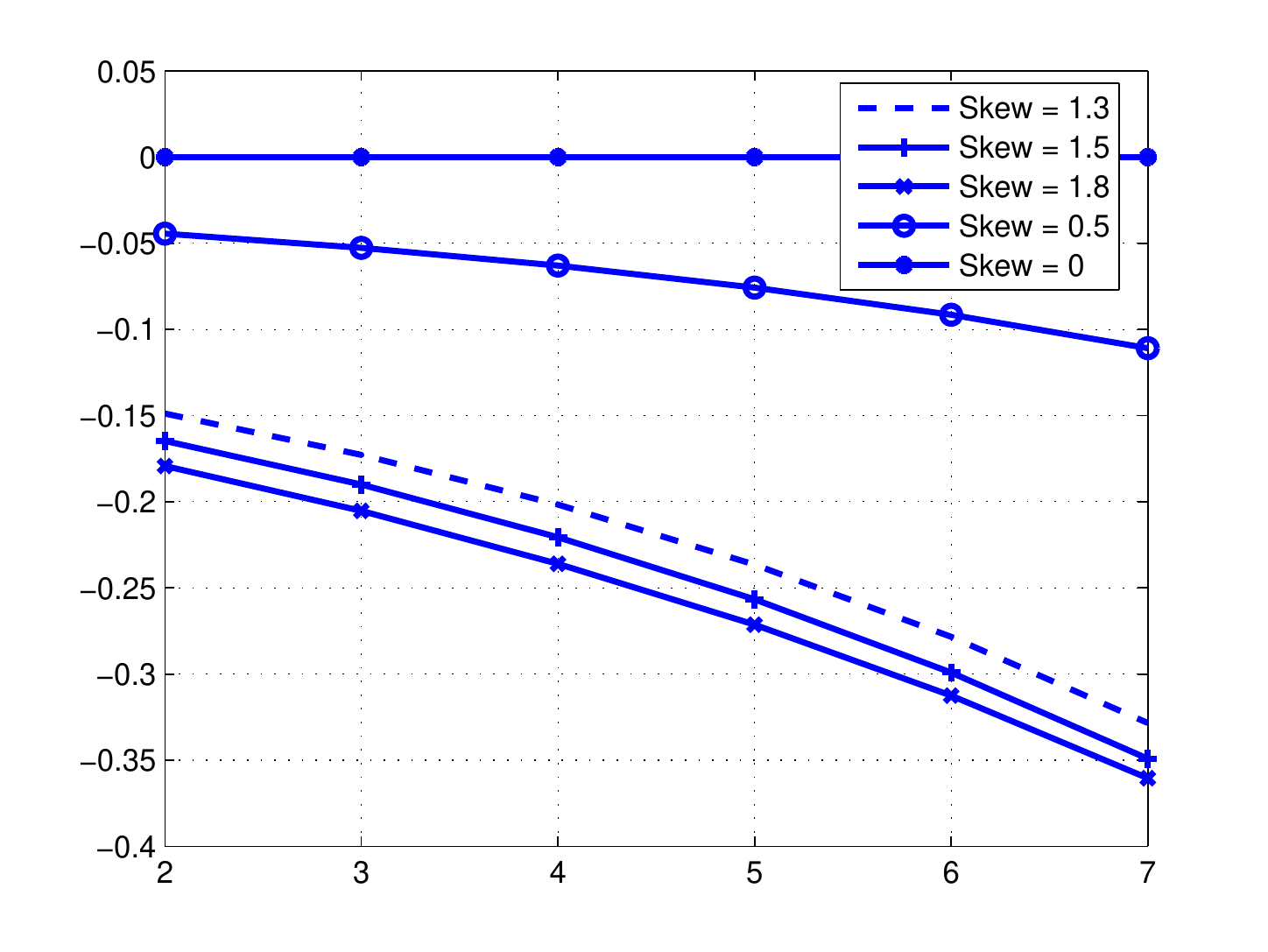}\\
  \caption{Decay factor for each skew value}\label{fig:df_skew_vs_snr}
\end{figure}

In Fig. (\ref{fig:df_skew_vs_snr}), we show the decay factors of Skewed-QPSK with $\sigma_{\Delta} = 0.1$[rads/symbol], for different skew values. We can see that for larger values of skew, the decay factor increases, and for a skew value of zero, i.e standard MPSK, we have a zero decay factor, which means that the decoding algorithms will not converge without pilots. Moreover, we observe that the decay factor increases in absolute value as the SNR increases. This is because the Tikhonov distributions in the forward and backward messages become narrower and thus the decay is accelerated.

\begin{figure}
  \centering
  \includegraphics[width=7.5cm]{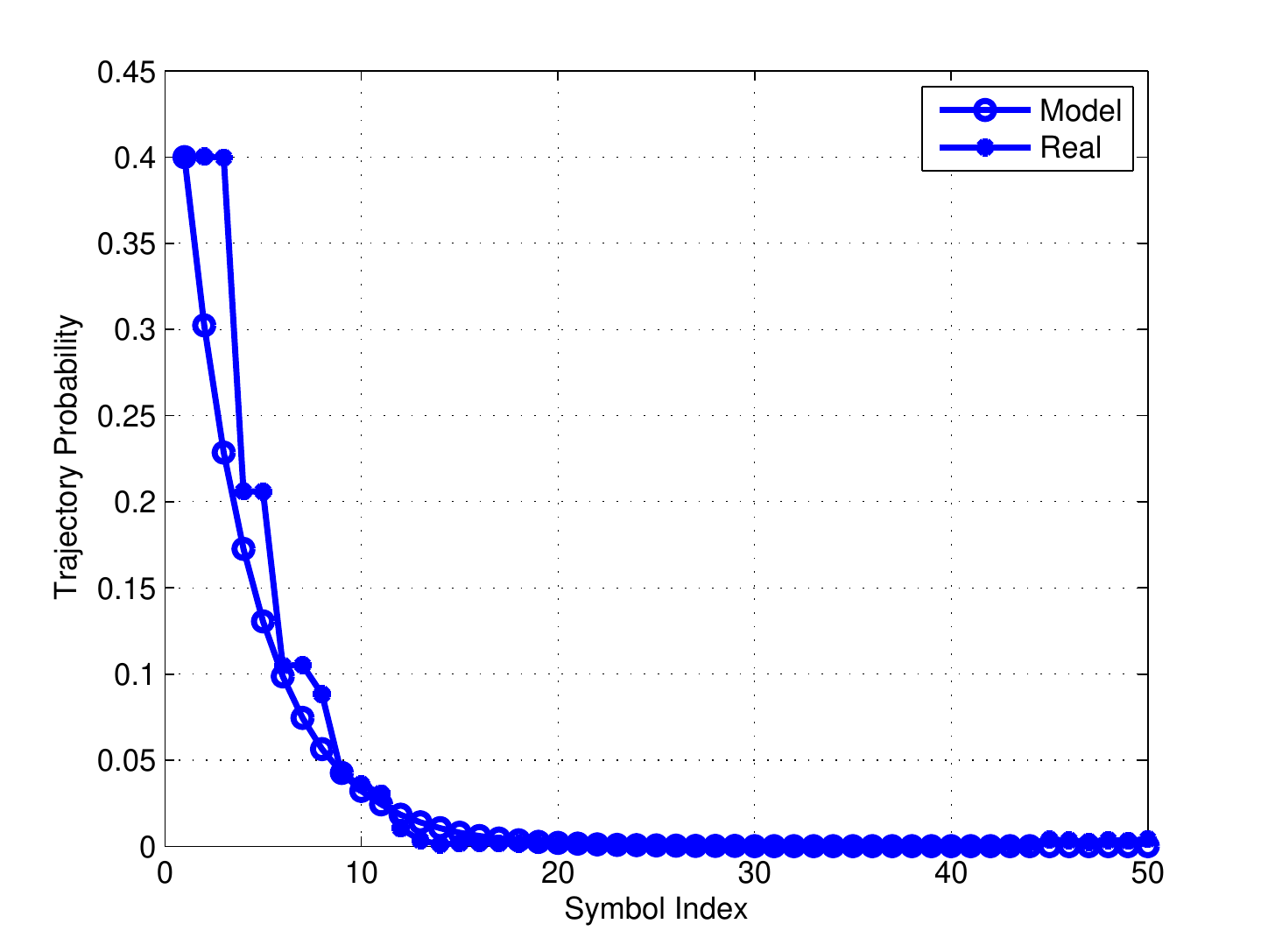}\\
  \caption{Simulation Vs Model}\label{fig:model_data}
\end{figure}

Next, we ran a Monte Carlo simulation of tracking two phase trajectories with $SNR = 4dB$ and $\sigma_{\Delta} = 0.1$[rads/symbol]. In Fig. (\ref{fig:model_data}), we show the mean value of the probability of the wrong trajectory ("Real") and the decay of the probability according to the decay factor ("Model").

\section{Simulation Results}
In this section we show the BER (Bit Error Rate) for the proposed scheme and compare it to an MPSK constellation with pilots. We have used a Monte Carlo simulation with QPSK and a skew value of $0.7$. We have also used a 4096 length, LDPC code with rate $0.89$ and for the QPSK case, we have used one pilot every 40 symbols and $\sigma_{\Delta}=0.2$ [rads/symbol]. The decoding algorithm chosen is the DP algorithm.

\begin{figure}
  \centering
  \includegraphics[width=7.5cm]{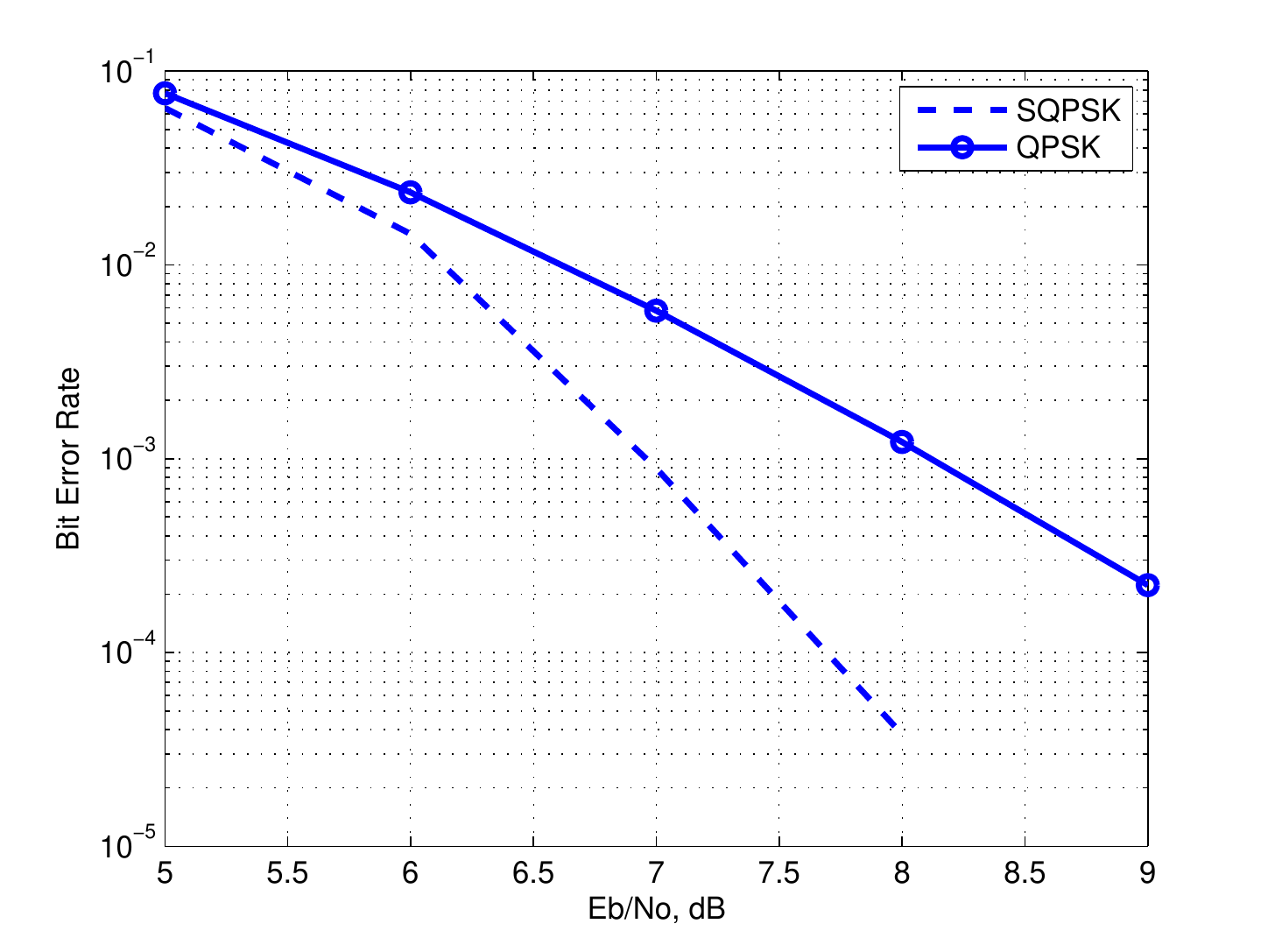}\\
  \caption{Bit Error Rate for SQPSK and QPSK}\label{fig:ber}
\end{figure}

The results in Fig. (\ref{fig:ber}), show that the proposed method is much superior to QPSK in very strong phase noise channels.

\section{Conclusions}
In this paper we proposed a new signal constellation for pilotless transmission of signals over Wiener phase noise channels. We also provided a method to analyze the performance of this constellation. This analysis can also be used to assess the performance of other signal constellations and provide insight.

\bibliographystyle{IEEEtrans}
\bibliography{strings}

\begin{thebibliography}{1}
\providecommand{\url}[1]{#1}
\csname url@samestyle\endcsname
\providecommand{\newblock}{\relax}
\providecommand{\bibinfo}[2]{#2}
\providecommand{\BIBentrySTDinterwordspacing}{\spaceskip=0pt\relax}
\providecommand{\BIBentryALTinterwordstretchfactor}{4}
\providecommand{\BIBentryALTinterwordspacing}{\spaceskip=\fontdimen2\font plus
\BIBentryALTinterwordstretchfactor\fontdimen3\font minus
  \fontdimen4\font\relax}
\providecommand{\BIBforeignlanguage}[2]{{%
\expandafter\ifx\csname l@#1\endcsname\relax
\typeout{** WARNING: IEEEtranS.bst: No hyphenation pattern has been}%
\typeout{** loaded for the language `#1'. Using the pattern for}%
\typeout{** the default language instead.}%
\else
\language=\csname l@#1\endcsname
\fi
#2}}
\providecommand{\BIBdecl}{\relax}
\BIBdecl

\bibitem{barb2011}
A.~Barbieri and G.~Colavolpe, ``On the information rate and repeat-accumulate
  code design for phase noise channels,'' \emph{IEEE Transactions on
  Communications}, vol.~59, pp. 3223 -- 3228, December 2011.

\bibitem{barb2005}
G.~Colavolpe, A.~Barbieri, and G.~Caire, ``Algorithms for iterative decoding in
  the presence of strong phase noise,'' \emph{IEEE Journal on Selected Areas in
  Communications}, vol.~23, pp. 1748 --1757, September 2005.

\bibitem{Hajimiri1998}
A.~Hajimiri and T.~H. Lee, ``A general theory of phase noise in electrical
  oscillators,'' \emph{IEEE Journal of Solid-State Circuits}, vol.~33, pp. 179
  -- 194, February 1998.

\bibitem{farbod2012}
F.~Kayhan and G.~Montorsi, ``Constellation design for channels affected by
  phase noise,'' \emph{CoRR}, vol. abs/1210.1752, 2012.

\bibitem{peleg2000iterative}
M.~Peleg, S.~Shamai, and S.~Galan, ``Iterative decoding for coded noncoherent
  mpsk communications over phase-noisy awgn channel,'' in \emph{Communications,
  IEE Proceedings-}, vol. 147, no.~2.\hskip 1em plus 0.5em minus 0.4em\relax
  IET, 2000, pp. 87--95.

\bibitem{shachar2012}
S.~Shayovitz and D.~Raphaeli, ``Multiple hypotheses iterative decoding of ldpc
  in the presence of strong phase noise,'' in \emph{Proceedings of The 2012
  IEEE 27th Convention of Electrical \& Electronics Engineers in Israel
  (IEEEI)}, 2012.

\end{thebibliography}

\end{document}